\documentclass[prb,showpacs]{revtex4}

\newcommand{\bu}[1]{{#1}^1B_{1u}}
\newcommand{\buminus}[1]{{#1}^1B_{1u}^-}
\newcommand{\ag}[1]{{#1}^1A_g}
\newcommand{\agplus}[1]{{#1}^1A_g^+}
\newcommand{\but}[1]{{#1}^3B_{1u}}
\newcommand{\butplus}[1]{{#1}^3B_{1u}^+}

\usepackage{graphicx}

\begin{document}

\widetext

\title{Relaxation energies and excited state structures of poly(para-phenylene)}

\author{Eric E. Moore$^{1*}$, William Barford$^{1,2**}$ and Robert J. Bursill$^3$}

\affiliation{
$^1$Department of Physics and Astronomy, University
of Sheffield, Sheffield, S3 7RH, United Kingdom\\
$^2$Cavendish Laboratory, University of Cambridge,
Cambridge, CB3 0HE, United Kingdom\\
$^3$School of Physics, University of New South Wales,\\
Sydney, New South Wales 2052, Australia
}

\begin{abstract}
We investigate the relaxation energies and excited state geometries of the light emitting polymer, poly(para-phenylene). We solve the Pariser-Parr-Pople-Peierls model using the density matrix renormalization group method.
We find that the lattice relaxation of the dipole-active  $\buminus{1}$ state is quite different from that of the $\butplus{1}$ state and the dipole-inactive $\agplus{2}$ state. In particular, the $\buminus{1}$ state is rather weakly coupled to the lattice and has a rather small relaxation energy $\sim 0.1$ eV. In contrast, the $\butplus{1}$ and $\agplus{2}$ states are strongly coupled with relaxation energies of $\sim 0.5$ and $\sim 1.0$ eV, respectively. 
By analogy to linear polyenes, we argue that this difference can be understood by the different kind of solitons present in the $\buminus{1}$, $\butplus{1}$ and $\agplus{2}$ states.
The difference in relaxation energies of the $\buminus{1}$ and $\butplus{1}$  states accounts for approximately one-third of the exchange gap in light-emitting polymers.

\end{abstract}


\maketitle

\section{Introduction}

Electron-lattice coupling has profound effects on the behavior of
conjugated polymers.  It is responsible for the self-trapping of
excited states and plays a vital role in determining the interconversion  between excited
states. Predicting  interconversion rates is important for understanding many electronic processes in conjugated polymers, \textit{e.g.}\ the determination of the singlet exciton yield in light emitting  polymers. Although not (directly) addressed here, the formation of
quinoid-like structures in poly(para-phenylene) is largely
responsible for the tendency of these molecules to planarize in the
excited state.

Electron-electron interaction is  also important in conjugated polymers. It is responsible for excitons with large binding energies and a large singlet-triplet exchange gap. The inter-play of both electron-lattice and electron-electron interactions leads to a complicated, but interesting  description of the excited states. In this paper we use the density matrix renormalization group (DMRG)\cite{white} method to solve the Pariser-Parr-Pople-Peierls model - a $\pi$-electron model that treats both electron-lattice and electron-electron interactions.

Poly(para-phenylene) has a $\agplus{1}$ ground state.  The lowest lying
optically allowed state is the $\buminus{1}$ state, and since this is the emitting
state in LEDs, its behavior is obviously important.  Another important state is the $\agplus{2}$ state. In light emitting polymers it is argued that this is the $m^1A_g$ state\cite{bursill2002}, which is strongly dipole connected to the $\buminus{1}$ state and consequently plays an important role in non-linear spectroscopy. As a consequence of both strong electronic correlations and electron-lattice coupling in linear polyenes, the relaxed energy of the $\agplus{2}$ state lies below the relaxed energy of the $\buminus{1}$ state, rendering these systems non-electroluminescent. We investigate the behavior of this state in poly(para-phenylene).
Because charge carriers are injected with arbitrary spin, triplet formation
is also of importance, so we also explore the behavior of the
$\butplus{1}$ state. 

The phenyl-based conjugated polymers are extrinsically semiconducting as a consequence of the chemical structure determined by the $\sigma$ bonds in the absence of $\pi$-conjugation. Thus, with all bond lengths equal there is still a semiconducting band gap. However, as for linear polyenes with extrinsic dimerization, coupling of the $\pi$-electrons to the lattice is still important, as it causes bond lengths to change resulting in exotic types of bound non-linear excitations. Understanding how the non-interacting description of the excited state structures change as a consequence of electronic interactions is a key goal of this work. To facilitate this goal, we map the excited state structures of poly(para-phenylene) into the more familiar description of mid-gap single-particle states and the associated solitons of linear polyenes.

Electron-lattice coupling in light emitting polymers has been investigated by a number of groups using a variety of methods. Beljonne \emph{et al.} studied the relaxation of the $1B_u$ singlet and triplet in short poly(paraphenylene vinylene) oligomers by solving intermediate neglect of differential overlap models using multi-reference configuration interactions. Although they found larger relaxation energies for the triplet than the singlet, the difference of ca.\ $0.1$ eV in the four-ring oligomer is considerably smaller than our results as presented in section IV.B.
Ambrosch-Draxl \emph{et al.}\cite{ambrosch} performed a density functional theory calculation on the ground state neutral structure of PPP, while Zojer \emph{et al.}\cite{zojer} performed  semi-empirical Austin Model 1 calculations on the ground state neutral and charged structures of PPP. Finally, Artacho \emph{et al.}\cite{artacho} performed a GWA-Bethe-Salpete equation calculation on the $\buminus{1}$ state of PPP.  We compare their predictions to ours in the results section. To our knowledge, this work presents the first large-scale calculation of the relaxation energies and geometrical structures of the $\buminus{1}$, $\butplus{1}$ and $\agplus{2}$ states in poly(para-phenylene) oligomers.

The experimental relaxation energies of the lowest-lying singlet exciton and doped polaron in a wide variety of polymers of different conjugation lengths has recently been presented by Wohlgenannt\cite{wohlgenannt}. Other experimental results are discussed in the results section, IV.B.

Extensive investigations of the excited states of poly(para-phenylene) for fixed geometries within the Pariser-Parr-Pople model, solved by the DMRG method, are described in ref\cite{bursill2002}.

In the next section we introduce the Pariser-Parr-Pople-Peierls model of conjugated polymers. Next, we solve the Pariser-Parr-Pople-Peierls model in the non-interacting limit, namely the Peierls model. We discuss soliton wavefunctions and soliton-antisoliton  confinement. Then we solve the full interacting model with the DMRG technique. We briefly discuss the DMRG algorithm before describing the results. Finally, we summarize and conclude.

\section{The Pariser-Parr-Pople-Peierls model}\label{Se:22}

The Pariser-Parr-Pople-Peierls model, $H_{PPPP}$, is a tight-binding model of the $\pi$-electrons that includes both long range Coulomb interactions and electron-lattice coupling. The electrons and lattice are coupled
together by the effects of changes in the bond lengths both on the
one-electron transfer integrals and the Coulomb interactions.
We treat these effects up to first order in the change
of bond length.   As the density-density correlator,
$(N_{i}-1)(N_j-1)$, decays rapidly with distance\cite{footnote1}, it is also a
reasonable  approximation to retain changes in the Coulomb
potential for only nearest neighbor interactions.

We thus define the Pariser-Parr-Pople-Peierls model as,
\begin{eqnarray}\label{Eq:7.1}
H_{PPPP} = && -2 \sum_i t_i \hat{T}_i + W \sum_i \Delta_i (N_{i+1}-1)(N_i-1)
\\
\nonumber && + \frac{1}{4\pi t \lambda} \sum_i \Delta_i^2 +
\Gamma\sum_i \Delta_i + U \sum_i \left(N_{i\uparrow}-
\frac{1}{2}\right) \left( N_{i\downarrow}- \frac{1}{2}\right)
\\
\nonumber &&  + \frac{1}{2} \sum_{i \ne j} V_{ij} (N_i-1)(N_j-1).
\end{eqnarray}

\begin{equation}\label{}
    \hat{T}_i = \frac{1}{2} \sum_{\sigma} \left( c_{i+1, \sigma}^{\dagger} c_{i, \sigma} +
    c_{i, \sigma}^{\dagger} c_{i+1 , \sigma} \right)
\end{equation}
is the bond-order operator, where $c_{i, \sigma}^{\dagger}$
creates an electron with spin $\sigma$ in the  $\pi$-orbital on
site $i$. The one-electron transfer integral is
\begin{equation}\label{Eq:7.3}
    t_i = t  + \frac{\Delta_i}{2},
\end{equation}
where
\begin{equation}\label{Eq:7.4}
   \Delta_i =  -2\alpha(u_{i+1} - u_{i}),
\end{equation}
and
$u_{i}$ is the displacement of the $i$th atom from its equilibrium position.
Thus, we define,
\begin{equation}\label{Eq:1}
    \delta u_i \equiv - \frac{\delta t_i}{\alpha} = - \frac{\Delta_i}{2\alpha},
\end{equation}
as the change in length of the $i$th bond from its initial - undistorted - value (determined by the $\sigma$ bonds).

The electron-phonon coupling constant is,
\begin{equation}\label{Eq:2.51}
   \lambda =  \frac{2 \alpha^2}{\pi K t}
\end{equation}
and $V_{ij}$ is the Ohno potential,
\begin{equation}\label{Eq:2.55}
    V_{ij} = \frac{U}{\sqrt{1+(Ur_{ij}/14.397)^2}},
\end{equation}
for the \textit{reference} (that is, the undistorted) structure,
where the bond lengths are in $\AA$.  We set
$U=10.06$ eV, $t=2.514$ eV, $\lambda = 0.12$, $\alpha = 4.67$ eV$\AA^{-2}$ and the undistorted bond length as $1.405$ $\AA$. 
The parameters used are essentially
those of ref\cite{bursill2002}, with $t$ and $\alpha$ chosen such that
if the bond lengths are those of ref\cite{bursill2002}, the transfer
integrals are as well, while $\lambda$ was chosen to obtain good
agreement with the ground state structure used in ref\cite{bursill2002}.

The second term on the right-hand-side of Eq.\ (\ref{Eq:7.1}) is
the change in the Coulomb interactions from changes in bond
length, where
\begin{equation}\label{Eq:7.6}
    W = \frac{1}{2\alpha}\left( \frac{\partial V_{ij}}{\partial \textbf{r}_{mn}}\right)_{{\textbf{r}_{ij} = r_0}}=
    \frac{U r_0 (U/14.397)^2}{2\alpha(1+(Ur_0/14.397)^2)^{3/2}}.
\end{equation}
It is instructive to
re-write this term as
\begin{equation}
-2\alpha W\sum_i (u_{i+1} - u_i)(N_{i+1}-1)(N_i-1),
\end{equation}
where we have used Eq.\ (\ref{Eq:7.4}). Expanding and resuming we
see that this term has two components. One  component is the
electron-phonon coupling arising from the change in the ionic
potentials,
\begin{equation}
-2\alpha W \sum_i(u_{i+1} - u_{i-1}) N_i.
\end{equation}
 The other component  represents the changes in
the nearest neighbor electron-electron interaction from the
change in bond length,
\begin{equation}
-2\alpha W \sum_i (u_{i+1} - u_i)N_{i+1}N_i.
\end{equation}
The first term on the right-hand-side of Eq.\ (\ref{Eq:7.1}) is
just the electron-phonon coupling arising from  the change in the
kinetic energy.

Notice that this model does not describe free rotations of phenyl rings relative to one another. Thus, its applicability is to ladder poly(para-phenylene), where the stereo-chemistry causes the rings to have a planar geometry, or polymers in the solid state, where ring rotations are more restricted.

As described in previous papers \cite{bursill1999, barford2001},
by using the Hellmann-Feynman theorem we  can derive a
self-consistent equation for $\Delta_i$ for any state,
\begin{equation}\label{Eq:7.10}
    \Delta_i = 2 \pi t \lambda \left( \langle \hat{T}_i \rangle - W \langle \hat{D}_i  \rangle - \Gamma \right),
\end{equation}
where
\begin{equation}\label{Eq:7.10a}
\hat{D}_i = (N_{i+1}-1)(N_i-1)
\end{equation}
 is the density-density
correlator for the $i$th bond.
$\Gamma$ is determined by imposing constant chain lengths,
$\sum_i \Delta_i =0$, implying that,
\begin{equation}\label{Eq:7.11}
   \Gamma =
   \overline{\langle \hat{T}_i \rangle} - W \overline{\langle \hat{D}_i \rangle},
\end{equation}
where the over-bar represents the spatial average.

\section{Non-interacting limit}\label{Se:NI}

In this section we describe the solutions of the Pariser-Parr-Pople-Peierls model in the non-interacting limit ($U=0$), namely the Peierls model.

\begin{figure}[tb]
\small
\begin{center}
\includegraphics[scale = 1.0]{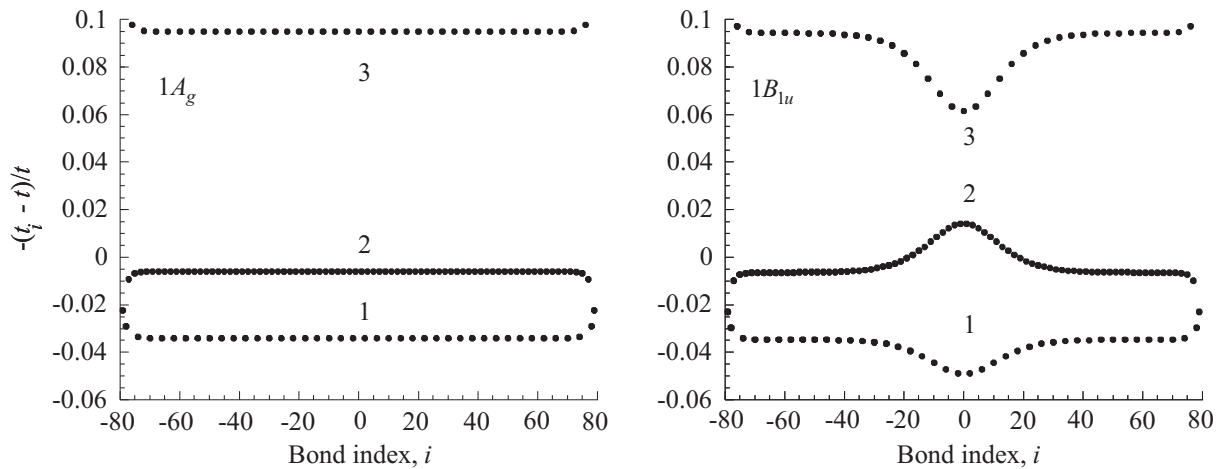}
\end{center}
\caption{The fractional change in transfer integrals of poly(para-phenylene) from the uniform
value, $t$, in the non-interacting limit. The relevant electronic states are indicated in the panels.  The labels refer to
the bonds shown in Fig.\ \ref{Fi:2}. Only the upper rung of bonds are
shown. Notice that the change in transfer integrals is opposite to the change in bond lengths.} \label{Fi:1}
\normalsize
\end{figure}

\begin{figure}[tb]
\begin{center}
\includegraphics[scale = 0.6]{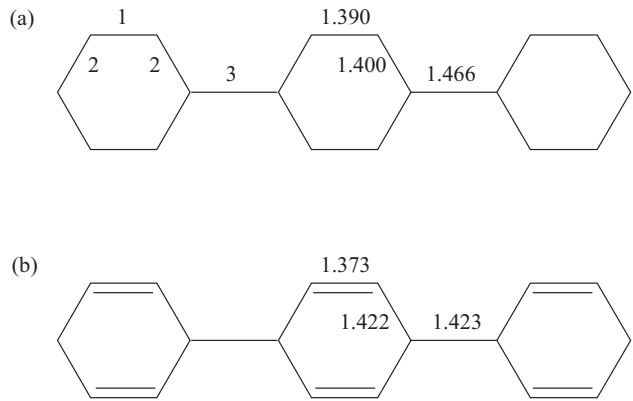}
\end{center}
\caption{(a) The bonds  illustrated in Fig.\ \ref{Fi:1} and Fig.\ \ref{Fi:5}, and bond lengths in $\AA$ of the ground state determined in the interacting limit.
(b) The quinoid structure of the $\buminus{1}$ state. Bond lengths in the center of the distortion in the interacting limit. (See section \ref{Se:1} and Fig. \ref{Fi:5})
} \label{Fi:2}
\end{figure}

Fig.\ \ref{Fi:1}  shows the fractional change in transfer integrals for the ground state, $\delta t_i$, defined in Eq.\ (\ref{Eq:1}). The bonds are defined by Fig.\ \ref{Fi:2}. Since the bonds are initially all of the same length, we see that the coupling of the $\pi$-electrons to the lattice has caused an effective `bond' alternation. The phenyl-ring bonds shorten  while the bridging bond lengthens. This is the benzoid structure, as the phenyl-ring bonds are roughly all of the same length.

To see this effective bond alternation  we define the summed bond distortions as,
\begin{equation}\label{}
    \delta t_n = \sum_{i \in \textrm{phenyl ring}}\delta t_i;\textrm{ odd }n
\end{equation}
and
\begin{equation}\label{}
    \delta t_n =\delta t_{i=\textrm{ bridging bond}};\textrm{ even }n.
\end{equation}
Then we defined the normalized, staggered and summed `bond' alternation, $\delta_n$, as,
\begin{equation}\label{Eq:12.11}
    \delta_n = \frac{\delta t_n}{t} (-1)^n.
\end{equation}

\begin{figure}[tb]
\small
\begin{center}
\includegraphics[scale = 0.6]{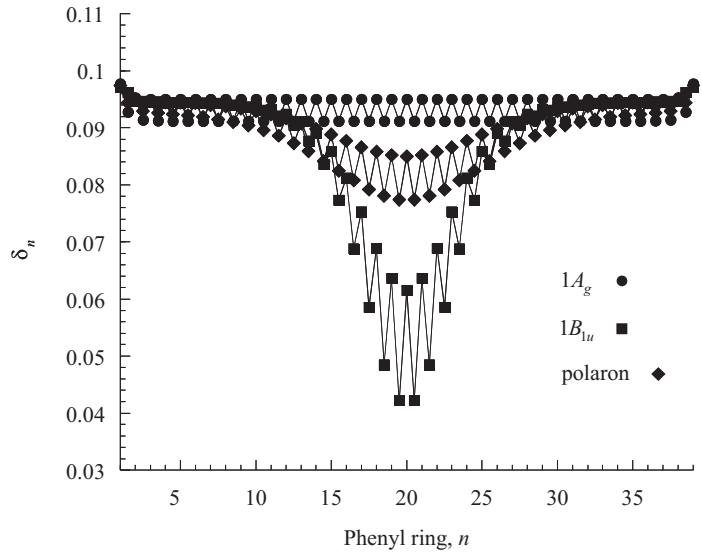}
\end{center}
\caption{The staggered, normalized and summed bond distortions of poly(para-phenylene) (as defined in Eq.\ (\ref{Eq:12.11})) in the non-interacting limit.} \label{Fi:4}
\normalsize
\end{figure}

Fig.\ \ref{Fi:4} shows $\delta_n$ for the ground state. Under this mapping the phenyl ring is equivalent to a double bond (or dimer) and the bridging bond is a single bond. As for polyenes, this effective alternation increases the semiconducting band gap. Note that end-effects coupled to the constraint of an overall constant contour length causes the oscillations in $\delta_n$: there are greater distortions in the phenyl rings at the end of the chain than in those in the middle of the chain. Thus, in the middle of the chain the summed distortion in bond lengths in a phenyl ring is not quite equal and opposite to the distortion of the bridging bonds.

Next consider the $1B_{1u}$ excited state structure, shown in Fig.\ \ref{Fi:1}. This is  the quinoid structure, illustrated in Fig.\ \ref{Fi:2} (b). In contrast to the ground state, there is now a significant variation in the bond lengths in the phenyl-ring: bonds labelled 1 shorten, while bonds labelled 2 lengthen. The bridging bond also shortens.

At first sight the excited state lattice distortions of poly(para-phenylene) represented in Fig.\ \ref{Fi:1} do not resemble that of a linear polyene. We therefore might enquire whether the geometrical defects (for example, solitons, polarons, \textit{etc.}) and their associated mid-gap electronic states also exist in an analogous manner in poly(para-phenylene). To show that bond defects do exist in an analogous manner to linear polyenes we again consider the summed bond distortions, defined by Eq.\ (\ref{Eq:12.11}).

The $1B_{1u}$ state structure is illustrated in this way in Fig.\ \ref{Fi:4}. The relaxed $1B_{1u}$ state creates a 
`polaronic' structure, whereby the average bond length in the phenyl ring increases while the bridging bond length decreases, but there is no reversal in bond distortions from the ground state. This polaronic structure of excited states occurs in extrinsically semiconducting polymers where the ground state is non-degenerate: reversing the sign of the bond distortions gives a higher energy\cite{brazovskii}. A bond defect, or soliton, separates two regions of opposite bond distortions. Creating  a soliton and antisoliton pair and moving them apart creates a region of reversed bonds. Thus, there is a linear confining potential between the soliton and antisoliton for large separations.
As in linear polyenes, these bond defects are also associated with mid-gap states.  

\begin{figure}[tb]
\begin{center}
\includegraphics[scale = 0.6]{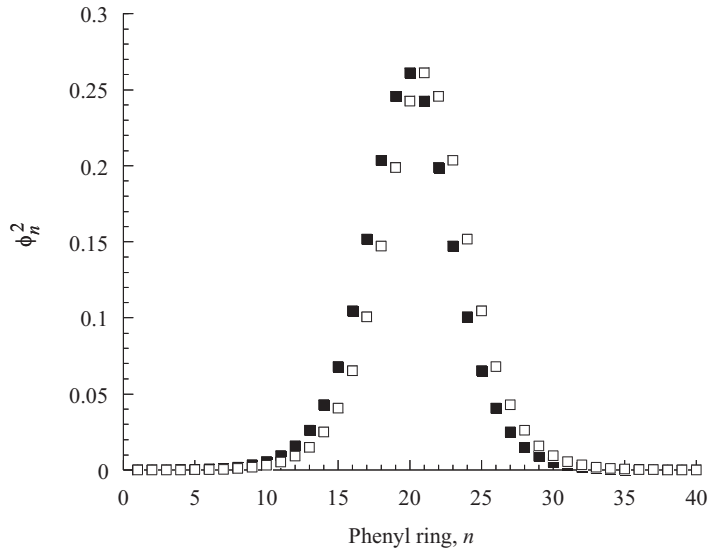}
\end{center}
\caption{The soliton (solid symbols) and anti-soliton (open symbols) probability densities (defined by Eq.\ (\ref{Eq:2})).
} \label{Fi:3}
\end{figure}

Associated with the two-mid gap single-particle states of the excited state are a bonding, $\psi_i^+$, and anti-bonding, $\psi_i^-$, molecular orbital (where $i$ is a site index). These molecular orbitals are analogous to the bonding and anti-bonding orbitals of molecular hydrogen\cite{ball}. The molecular orbitals are constructed from localized Wannier functions, $\phi_i$ and $\bar{\phi}_i$, which represent the soliton and anti-soliton respectively. In particular,
\begin{equation}\label{Eq:18}
    \psi_i^{\pm} = \frac{1}{\sqrt{2}} ( \phi_i \pm \bar{\phi}_i),
\end{equation}
or inverting,
\begin{equation}\label{}
    \phi_i = \frac{1}{\sqrt{2}} ( \psi_i^+ + \psi_i^-)
\end{equation}
and
\begin{equation}\label{}
    \bar{\phi}_i = \frac{1}{\sqrt{2}} ( \psi_i^+ - \psi_i^-).
\end{equation}
In linear polyenes with degenerate ground states the soliton and antisoliton are widely separated. However, as described above, they are confined in extrinsic semiconductors. This confinement is illustrated for poly(para-phenylene) in Fig.\ \ref{Fi:3}, which shows the soliton and antisoliton probability density summed over each phenyl ring,
\begin{equation}\label{Eq:2}
    \phi^2_n = \sum_{i \in \textrm{phenyl ring}} \phi_i^2.
\end{equation}
We see that the soliton and anti-soliton wavefunctions are centered on neighboring phenyl rings in the middle of the chain.

We now use this description of the molecular orbital defect states (Eq.\ \ref{Eq:18}) to describe the excited states.
As first shown by Ball \textit{et al.} for \textit{trans}-polyacetylene\cite{ball}, as a consequence of spin and spatial symmetries the relaxed $\buminus{1}$ and $\butplus{1}$ states have quite different solitonic characteristics. These differences become important when the spin degeneracy is lifted by electronic interactions, and they help explain the quite different geometrical distortions of these two states in the interacting limit. 

We first review the argument of ref\cite{ball}.
First, let us consider the singlet, $1^1B_{1u}$ state.  We write this as,
\begin{equation}\label{Eq:4.19}
    |1^1B_{1u} \rangle = \frac{1}{\sqrt{2}} \left( c_{+ \uparrow}^{\dagger}c_{- \downarrow}^{\dagger} -
    c_{+ \downarrow}^{\dagger}c_{- \uparrow}^{\dagger} \right) |V \rangle,
\end{equation}
where $|V \rangle$ represents the occupied sea of valence states and $c_{\pm \sigma}^{\dagger}$ creates an
 electron with spin $\sigma$ in the mid gap state $|\psi^{\pm}\rangle$. 
If $c_{\sigma}^{\dagger}$ and $\bar{c}_{\sigma}^{\dagger}$ creates an electron in the states
 $|\phi\rangle$ and $|{\bar \phi} \rangle$, respectively, then
\begin{equation}\label{Eq:4.20}
    c_{\pm \sigma}^{\dagger} = \frac{1}{\sqrt{2}} \left( c_{\sigma}^{\dagger} \pm
     \overline{c}_{\sigma}^{\dagger} \right).
\end{equation}
Inserting Eq.\ (\ref{Eq:4.20}) into Eq.\ (\ref{Eq:4.19}), we have
\begin{equation}\label{Eq:4.21}
    |1^1B_{1u} \rangle = \frac{1}{\sqrt{2}} \left( c_{ \uparrow}^{\dagger}c_{ \downarrow}^{\dagger} -
 \overline{c}_{ \uparrow}^{\dagger}\overline{c}_{ \downarrow}^{\dagger} \right) |V \rangle.
\end{equation}
$c_{ \uparrow}^{\dagger}c_{ \downarrow}^{\dagger}$ creates a pair of electrons in the soliton, so it is negatively
charged and spinless, while the anti-soliton contains no electrons, so it is positively charged and also spinless.
 Similarly, $\overline{c}_{ \uparrow}^{\dagger}\overline{c}_{ \downarrow}^{\dagger}$ creates a pair of
 electrons in the anti-soliton, while the soliton contains no electrons.
The $ 1^1B_{1u}$ state is therefore a linear superposition of spinless positively and negatively charged
 soliton-antisoliton pairs.

A similar argument applies to the triplet,  $1^3B_{1u}$ state,
\begin{eqnarray}\label{Eq:4.22}
    |1^3B_{1u} \rangle && = \frac{1}{\sqrt{2}} \left( c_{+ \uparrow}^{\dagger}c_{- \downarrow}^{\dagger} +
    c_{+ \downarrow}^{\dagger}c_{- \uparrow}^{\dagger} \right) |V \rangle \nonumber \\
 && = \frac{1}{\sqrt{2}} \left( c_{\uparrow}^{\dagger}\overline{c}_{\downarrow}^{\dagger} +
    c_{ \downarrow}^{\dagger}\overline{c}_{ \uparrow}^{\dagger} \right) |V \rangle,
\end{eqnarray}
showing that it is a linear superposition of neutral
spin-$\frac{1}{2}$ soliton-antisoliton pairs.

Electronic interactions have a significant affect on these states. The spinless oppositely charged solitons of the $\bu{1}$ state bind to form an exciton-polaron. Conversely, the neutral spin-$1/2$ solitons of the $\but{1}$ state do not bind, but cause a locally strong lattice distortion. We investigate these structures in the next section.

To aid in our understand of exciton-polaron structures (to be described below) and relaxation energies, we also investigate charged (polaron) states in the non-interacting limit.  Fig.\ \ref{Fi:4} shows the polaronic structure associated with a doped particle. Table \ref{Ta:2} lists the relaxation energies of the $1B_{1u}$, $2A_g$ and charged (polaron) states for different oligomer lengths. We note that the relaxation energy of the $1B_{1u}$ state is considerably greater than for the polaron, and that the relaxation energies reduce as the oligomer lengths increase.

\begin{table}[tp]
\small \caption{The relaxation energies of the $1B_{1u}$, $2A_g$ and charged (polaron) states  for para-phenylene oligomers (in eV) calculated from the Peierls model.}
\begin{center}
\begin{tabular}{cccc}
Number of phenyl rings   & $1B_{1u}$  & $2A_{g}$  & polaron \\
\hline
$4$ & $0.23$ & $0.14$ & $0.06$ \\
$8$ & $0.15$ & $0.10$ & $0.04$ \\
$20$ & $0.08$ & $0.06$ &  $0.02$ \\
$40$ & $0.06$ & $0.03$ & $0.01$ \\
\hline
\end{tabular}
\normalsize
\end{center}
\label{Ta:2}
\end{table}

In general, poly(para-phenylene) is not planar because of the steric repulsion of the hydrogen atoms on neighboring phenyl rings. The torsional angle between adjacent phenyl rings for a single chain is estimated to be $27^0$ (ref\cite{ambrosch}) and $34^0$ (ref\cite{artacho}). Packing in a crystalline environment planarizes the chain, and in this case the torsional angle is estimated to be $17^0$ (ref\cite{ambrosch}). The quinoid structure of the excited state also planarizes the chain, because in this structure the bridging bond has more double bond character, and thus twisting the rings reduces the bond integral and hence increases the energy more than in the benzoid structure. The torsional angle in the middle of the distortion is estimated to reduce to $\sim 9^0$ (ref\cite{artacho}). As discussed in section \ref{Se:22}, we do not model bond rotations in this work.

\section{Interacting limit}

It is well established that electron-electron interactions enhance the bond alternation in the ground state\cite{horsch} and generally enhance the size of the lattice distortions for excited states\cite{bursill1999, barford2001} of linear polyenes. The enhancement is greater and the electron-lattice relaxation energy is larger for states with covalent character relative to states that are entirely ionic in character. Thus, the $\butplus{1}$ and $\agplus{2}$ states undergo a greater electron-lattice relaxation than the $\buminus{1}$ state. These features also occur for the electronic states of light emitting polymers, as we  describe in this section. First we describe the DMRG algorithm for solving the  Pariser-Parr-Pople-Peierls model.

\subsection{Density Matrix Renormalization Group}

The DMRG method used is basically that of ref\cite{bursill2002},
although instead of a three-block method, a four-block method is used,
with two ``middle'' blocks being inserted into the middle of the chain
each time the chain is grown. 

The procedure is as follows:
\begin{itemize}
\item{
Initial, exact blocks (spanning the full Hilbert space) are constructed containing three carbon sites (or $\pi$-orbitals) each.  These are augmented in the 4-block DMRG program, which creates the biphenyl structure, to
construct optimized blocks for a phenyl ring.}
\item{These blocks are
further optimized by constructing a four-ring system, with the middle two
blocks being obtained from a biphenyl calculation with periodic
boundary conditions applied. End and middle blocks are reoptimized in-situ by performing single-block rotations and truncations.}  
\item{
The optimized middle and end phenyl ring basis states are then used
in the infinite lattice algorithm to grow larger oligomers, with two
middle blocks being inserted with each iteration and augmented with
the end blocks from the last iteration to provide the end blocks for
the next iteration.}
\item{At the target chain size a finite-lattice sweep is performed.}
\end{itemize}

\subsubsection{Convergence}

\begin{figure}[tb]
\small
\begin{center}
\includegraphics[width = 3in]{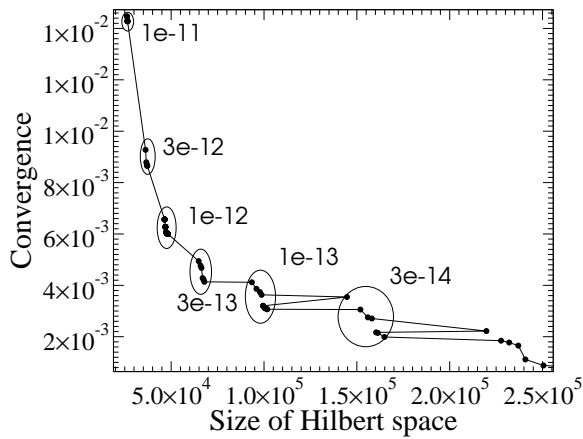}
\end{center}
\caption{The ground state energy (relative to the lowest calculated
  energy) plotted against Hilbert space size for a number of different
  values of $m$ and $\epsilon$.  Groups of data points with the same
  value for $\epsilon$ are labelled.}\label{Fi:7}
\normalsize
\end{figure}

The DMRG convergence parameters used were determined by calculating
the ground state energy for a range of different values, the results
of which are shown in Fig.\ \ref{Fi:7}.  
There are two truncation parameters in the calculation. $m$ is the truncation parameter for the spin and charge sector of a given block with largest number of states. This controls the overall truncation of a block. The superblock Hilbert space is controlled by $\epsilon$, the product of the density matrix eigenvalues for each block state comprising a superblock state.
The size of the Hilbert space is
largely independent of $m$ and the speed of the calculation is largely
determined by the size of the Hilbert space.  Since for a given
$\epsilon$ increasing $m$ yields improved results, but has little
effect on the speed of the calculation, $m$ was fixed at 50. The
results at $\epsilon = 3 \times 10^{-12}$ are within approximately
$0.01$ eV of the lowest energy values, so can be taken to be
converged to approximately that order of magnitude, and this value was
used for all the calculations reported.

\subsection{Results}\label{Se:1}

\begin{table}[tp]
\small \caption{The vertical and relaxation energies of para-phenylene oligomers (in eV) calculated from the Pariser-Parr-Pople-Peierls model. }
\begin{center}
\begin{tabular}{ccccc}
{State}   & \multicolumn{2}{c} \textit{Vertical transition energy}& \multicolumn{2}{c} \textit{Relaxation energy}\\
\hline
      &  $N=4$ & $N=8$  & $N=4$ & $N=8$ \\
\hline
$1^3B_{1u}^+$  & $3.29$ & $3.17$ & $0.58$ & $0.41$ \\
$1^1B_{1u}^-$  & $4.21$ & $3.96$ & $0.17$ & $0.06$  \\
$2^1A_g^+$  &  $5.52$ & $5.26$  & $1.28$ & $1.14$ \\
\hline
\end{tabular}
\normalsize
\end{center}
\label{Ta:1}
\end{table}

\begin{figure}[tb]
\small
\begin{center}
\includegraphics[scale = 1.]{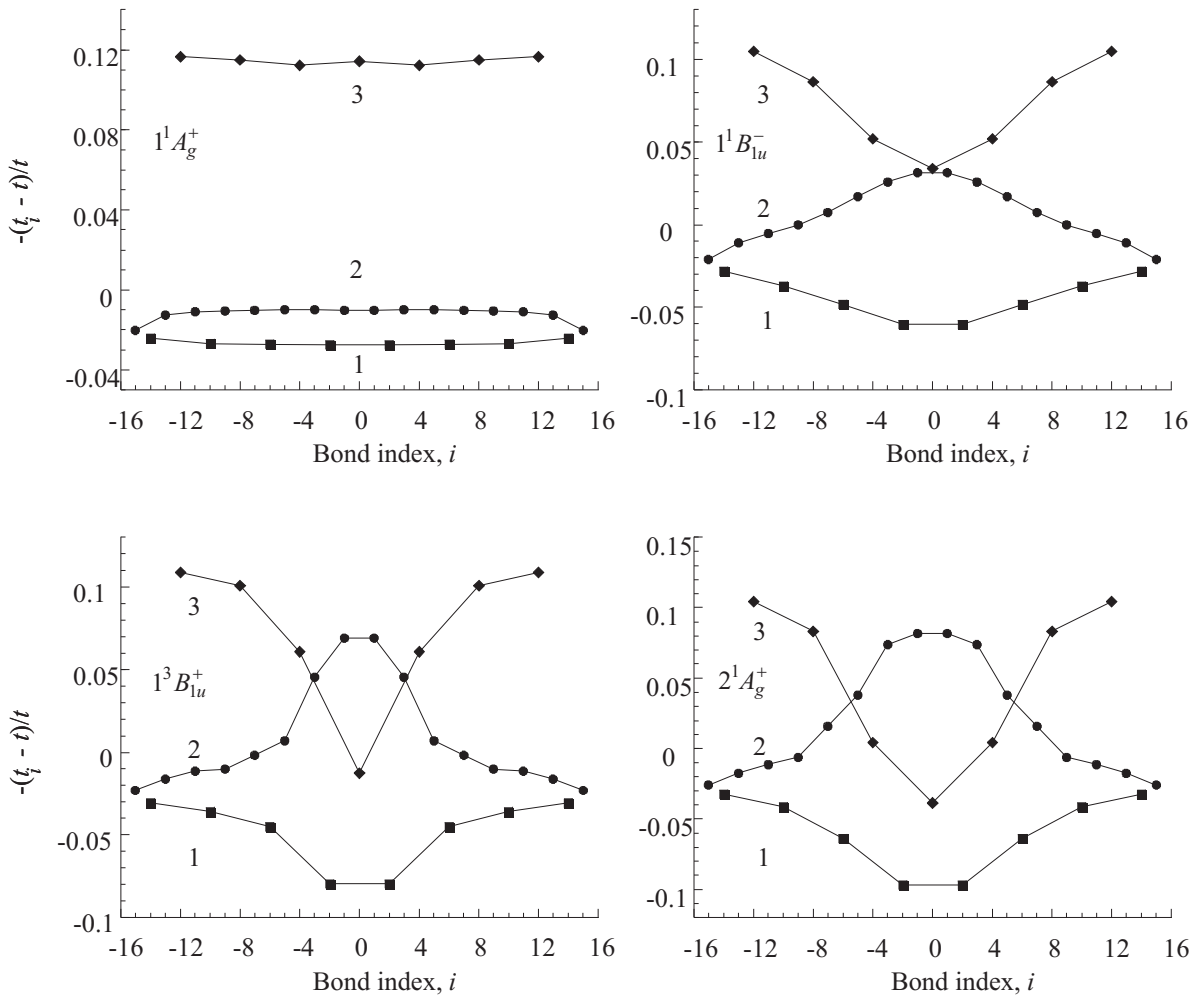}
\end{center}
\caption{The fractional change in transfer integrals of eight-ring (para-phenylene) oligomers from the uniform
value, $t$, in the interacting limit. The labels refer to
the bonds shown in Fig.\ \ref{Fi:2}. Only the upper rung of bonds are
shown.} \label{Fi:5}
\normalsize
\end{figure}

Table \ref{Ta:1} lists the vertical and relaxed energies of the $\butplus{1}$, $\buminus{1}$ and $\agplus{2}$ states for $4$ and $8$ ring para-phenylene oligomers. As in linear polyenes\cite{barford2001, barford2005}, the relaxation energy of the $\buminus{1}$ state is small, whereas the relaxation energy of the $\butplus{1}$ state is large.
The experimentally determined relaxation energy of the $\buminus{1}$ state in the related polymer poly(para-phenylene vinylene) has been reported as $0.07$ in ref\cite{liess}. We may also deduce the relaxation energy in poly(para-phenylene) and ladder poly(para-phenylene) from Fig.\ 3 of ref\cite{hertel} by noting that the ratio of the intensities of the $0-1$ to $0-0$ vibronic peaks in the absorption or emission spectra is $S$, the effective Huang-Rhys factor. The relaxation energy is then $\hbar \omega \times S$, where $\hbar \omega$ is the characteristic phonon frequency $\sim 0.2$ eV. Thus, using $S = 0.6$ for ladder poly(para-phenylene), $S = 1.2$ for poly(para-phenylene) and $\hbar \omega = 0.2$ eV gives relaxation energies of $0.12$ eV and $0.24$ eV for ladder poly(para-phenylene) and  poly(para-phenylene), respectively. The larger relaxation energy for poly(para-phenylene) is expected, as the rings are free to rotate, and this result is consistent with a
 calculated value of $0.22$ eV reported in ref\cite{artacho}.
The relaxation energy of the $\buminus{1}$ state in the interacting limit is intermediate between the relaxation energy of the $1B_{1u}$ state and polaron in the non-interacting limit, as listed in Table \ref{Ta:2}. This illustrates the exciton-polaron nature of the $\buminus{1}$ state, as further discussed below.

The relaxation energy of the $\agplus{2}$ state is also large, but not large enough to cause an energy level reversal of the $\buminus{1}$ and $\agplus{2}$ states.
The difference in relaxation energies between the $\butplus{1}$ and $\buminus{1}$ states increases the $0-0$ energy singlet-triplet exchange gap from the vertical gap of $\sim 0.6 $ eV to $\sim 0.9$ eV, in good agreement with experiment\cite{kohler}.
 We also see that the relaxation energy reduces with chain size, consistent with an increased delocalization of the excitations and consequently a diminished effective electron-lattice coupling, and in agreement with a wide number of experimental results\cite{wohlgenannt}.

\begin{figure}[tb]
\small
\begin{center}
\includegraphics[scale = 0.6]{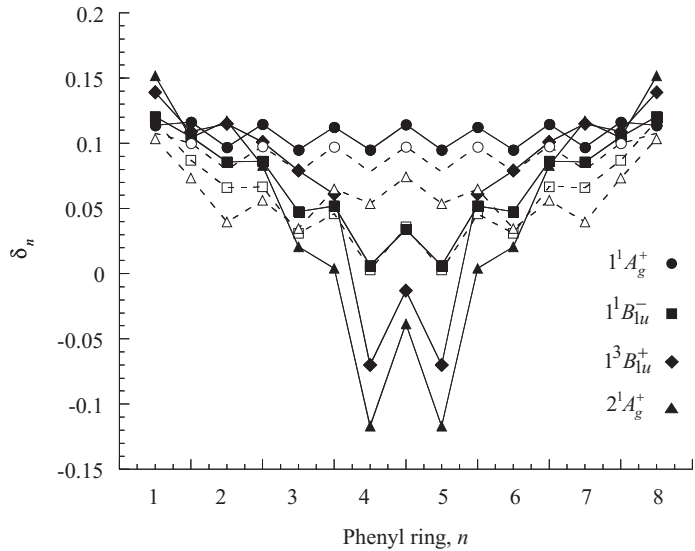}
\end{center}
\caption{The staggered, normalized and summed bond distortions of eight-ring para-phenylene oligomers (as defined in Eq.\ (\ref{Eq:12.11})) in the interacting limit (filled symbols and solid lines) and non-interacting limit (empty symbols and dashed lines). Note that in the non-interacting limit the $1^1B_{1u}^-$ and $1^3B_{1u}^+$ states are equivalent.} \label{Fi:6}
\normalsize
\end{figure}

Next, we consider the associated geometrical structures. These are plotted in Fig.\ \ref{Fi:5} for the normalized changes in transfer integrals and in Fig.\ \ref{Fi:6} for the staggered, summed bond distortions. As predicted, the ground state alternation is enhanced in the interacting limit over the non-interacting limit by $8\%$. The  bond lengths, calculated using Eq.\ (\ref{Eq:1}), are shown in Fig.\ \ref{Fi:2}.  

The $1^1B_{1u}^-$ state is now an exciton-polaron\cite{grabowski}. Its structure is qualitatively similar in both the non-interacting and interacting limits, as the soliton-antisoliton confinement due to linear confinement arising from the effective extrinsic bond alternation has a rather similar effect to electron-hole attraction. However, as already predicted, the $1^3B_{1u}^+$ state has a more pronounced distortion because it has some covalent character. Indeed, there is a change of sign in the effective bond alternation. The middle bridging bond becomes a `short' bond, while the adjacent phenyl-ring become `long' bonds. Similarly, the $\agplus{2}$ state shows a significant structural distortion, with a change of sign of the bond alternation. The lattice distortions of the $\buminus{1}$, $\butplus{1}$ and $\agplus{2}$ states - as defined by the summed bond distortions of Eq.\ (\ref{Eq:12.11}) and shown in Fig.\ \ref{Fi:6} - are qualitatively similar to those of linear polyenes with extrinsic dimerizarion\cite{barford2005}.

The different relaxation energies and geometrical structures of the singlet and triplet $B_{1u}$ states in the interacting limit is obviously related to the different kind of solitons comprising these states, as described in section \ref{Se:NI}. In particular, the electronic interactions induce a strong coupling of the neutral soliton to the bond-order correlation, causing a significant distortion for the triplet state. In contrast, the charged solitons weakly couple to the bond-order correlation, and thus the singlet state is more weakly coupled to the lattice. Since the $\agplus{2}$ state has an admixture of charged and neutral solitons, it also couples more strongly to the lattice than the 
$1^1B_{1u}^-$ state.

\section{Concluding remarks}

Understanding electron-lattice relaxation, self-trapping and calculating Huang-Rhys factors has important implications for predicting electronic processes in conjugated polymers, for example, exciton migration, recombination and inter-conversion mechanisms. By solving the Pariser-Parr-Pople-Peierls model of conjugated polymers for poly(para-phenylene) by the DMRG method we have shown that the lattice relaxation of the $\buminus{1}$ state is quite different from that of the $\butplus{1}$ and $\ag{2}$ states. In particular, the $\buminus{1}$ state is rather weakly coupled to the lattice and has a rather small relaxation energy $\sim 0.1$ eV. In contrast, the $\butplus{1}$ and $\ag{2}$ states are strongly coupled with relaxation energies of $\sim 0.5$ and $\sim 1.0$ eV, respectively. 
By analogy to linear polyenes, we argue that this difference can be understood by the different kind of solitons present in the $\buminus{1}$, $\butplus{1}$ and $\agplus{2}$ states. The difference in relaxation energies of the $\buminus{1}$ and $\butplus{1}$  states accounts for approximately one-third of the exchange gap in light-emitting polymers.

The results of this calculation present a number of questions. First, are vertical and relaxed states solvated by the environment by the same amount ? If not, then the energy gap between the vertical and relaxed states for a polymer in the solid state will be different from that predicted here. Second, is the large relaxation energy of the $\agplus{2}$ state evident experimentally ? Finally, are the different relaxation energies of the $\buminus{1}$ and $\butplus{1}$ states relevant to the issue of the singlet-triplet exciton fraction in light emitting polymers\cite{shuai, tandon, beljonne2, barford2004} ?

\begin{acknowledgements}
This work was supported by the EPSRC (U.K.) (GR/R02177/01 and GR/R03921/01).
W.\ B.\ gratefully acknowledges the financial support of the Leverhulme
Trust, and thanks the Cavendish Laboratory and Clare Hall,
Cambridge for their hospitality. R.\ J.\ B.\ was supported by the Australian Research Council and the J.\ G.\ Russell Foundation.
\end{acknowledgements}

\end{document}